\documentclass[floats,aps,amsmath,showpacs,twocolumn]{revtex4}

\usepackage{graphicx}

\begin{document}

\newcommand{\QATOP}[2]{#1\atop #2}

\title{Semiclassical Prediction for Shot Noise in Chaotic Cavities}
\date{\today }
\author{Petr Braun$^{1,2}$, Stefan Heusler$^1$, 
Sebastian M{\"u}ller$^1$,
Fritz Haake$^1$}

\address{$^1$Fachbereich Physik, Universit{\"a}t Duisburg-Essen,
45117 Essen, Germany\\
$^2$Institute of Physics, Saint-Petersburg University, 198504
Saint-Petersburg, Russia}

\begin{abstract}
  We show that in clean chaotic cavities the power of shot noise takes
  a universal form. Our predictions go beyond previous results from
  random-matrix theory, in covering the experimentally relevant case
  of few channels. Following a semiclassical approach we evaluate the
  contributions of quadruplets of classical trajectories to shot
  noise.  Our approach can be extended to a variety of transport
  phenomena as illustrated for the crossover between symmetry classes
  in the presence of a weak magnetic field.
\end{abstract}

\maketitle \date{\today}

%73.23.-b Electronic transport in mesoscopic systems
%72.20.My Galvanomagnetic and other magnetotransport effects
%72.15.Rn Localization effects (Anderson or weak localization)
%05.45.Mt Quantum chaos; semiclassical methods
%03.65.Sq Semiclassical theories and applications

%73.20.Fz Weak or Anderson localization (surface/interface states)
%05.40.Ca Noise
%03.65.Nk Scattering theory
%73.63.-b Electronic transport in nanoscale materials and structures
%(see also 73.23.-b Electronic transport in mesoscopic systems)
%73.23.Ad Ballistic transport (see also 75.47.Jn Ballistic magnetoresistance
%in magnetic properties and materials)
%75.47.-m Magnetotransport phenomena; materials for magnetotransport
%(for spintronics, see 85.75.-d; see also 72.15.Gd, 73.50.Jt, 73.43.Qt,
%and 72.25.-b in transport phenomena)

Ballistic chaotic cavities have universal transport properties, just
as do disordered conductors. The explanation of such universality
cannot rely on any disorder average but must make do with chaos in an
{\it individual} clean cavity. We shall present here the semiclassical
explanation of shot noise, relating the
quantum properties of chaotic cavities to the interference between
contributions of mutually close classical trajectories.  Similar
methods have recently been used for explaining universal spectral
fluctuations of chaotic quantum systems \cite{SR,EssenFF}, and to
calculate the universal mean conductance in \cite{RS,EssenCond}.

Following Landauer and B{\"u}ttiker \cite{Landauer,Buettiker}, we
treat transport as scattering between two leads attached to the
cavity. One lead is assumed to support $N_1$ ingoing channels and the
second one $N_2$ outgoing channels. In contrast to the random-matrix
treatment of \cite{Buettiker,Beenakker} and work on quantum graphs in
\cite{Schanz}, our results cover all orders in the inverse number of
channels, $N=N_1+N_2$, and thus apply to the experimentally relevant
case of few channels \cite{BeeHout}. Previously unknown and
surprisingly simple expressions for the shot noise arise, both with
and without time reversal invariance (see Eq.~(\ref{univshot}) below).

The transition amplitudes between ingoing channels $a_1$ and outgoing
channels $ a_2 $ define an $N_1\times N_2$ matrix $t=\{t_{a_1a_2}\}$.
That matrix
determines the power of shot noise as $P=\langle\mathrm{tr}%
(tt^\dagger-tt^\dagger tt^\dagger)\rangle$, in units
$\frac{2e^3|V|}{\pi\hbar }$ depending on the voltage $V$; for us,
$\langle\ldots\rangle$ denotes an average over a small energy
interval. Previous work had involved averages over ensembles of
matrices $t$ and obtained \cite{Buettiker,Beenakker}
\begin{equation}  \label{RMT}
P=\frac{N_1^2N_2^2}{N^3}+\Big(\frac{2}{\beta}-1\Big)
\frac{N_1N_2(N_1-N_2)^2}{N^4}
+{\cal O}\Big(\frac{1}{N}\Big)\,;
\end{equation}
here $\beta=1$ refers to the so-called ``orthogonal case" of
time-reversal invariant dynamics; if a magnetic field is applied to
break time-reversal invariance (``unitary case", $\beta=2$), the
second (``weak localization'') term disappears. Higher
orders in $\frac{1}{N}$ are as yet unknown.

In the semiclassical limit, each transition amplitude $t_{a_{1}a_{2}}$
is given by a sum over trajectories $\alpha $ leading from an ingoing
channel $a_{1}$ to an outgoing channel $a_{2}$, $t_{a_{1}a_{2}}\sim
\sum_{\alpha(a_{1}\to a_{2})}\frac{A_{\alpha
  }}{\sqrt{T_{H}}}\mathrm{e} ^{{\rm i} S_{\alpha }/\hbar }$
\cite{Richter}. It can be shown that the absolute value of the initial
angle of the relevant trajectories (i.e.  the angle enclosed between
the initial piece and the direction of the lead) is dictated by the
ingoing channel, whereas the final angle is determined by the outgoing
channel. The contribution of each trajectory depends on the Heisenberg
time $T_{H}=\frac{\Omega}{(2\pi\hbar)^{f-1}}$, with $\Omega$ the
volume of the energy shell and $f$ the number of freedoms. The factor
$A_{\alpha }$ is determined by the stability of the trajectory, and
the phase is proportional to the classical action $S_{\alpha }$.

With the transition amplitudes thus semiclassically approximated, the
quadratic term $\langle \mathrm{tr}(tt^{\dagger })\rangle$ becomes a
double sum over trajectories. That double sum, which
actually is the mean conductance, was evaluated in
\cite{EssenCond} as $\frac{N_{1}N_{2}}{N-1+2/\beta}$.
The quartic contribution to shot noise
turns into a sum over quadruplets of trajectories
\begin{eqnarray}
\!\!\!\!\!\!\!\!\!\!\!\! &&\langle \mathrm{tr}(tt^{\dagger }
tt^{\dagger })\rangle
=\sum_{\QATOP{a_{1},c_{1}}{a_{2},c_{2}}}t_{a_{1}a_{2}}t_{c_{1}a_{2}}^{\ast
}t_{c_{1}c_{2}}t_{a_{1}c_{2}}^{\ast }  \nonumber  \label{quadsum} \\
\!\!\!\!\!\!\!\!\!\!\!\! &&=\frac{1}{T_{H}^{2}}
\left\langle \sum_{\QATOP{a_{1},c_{1}}{%
a_{2},c_{2}}}\sum_{\alpha ,\beta ,\gamma ,\delta }\!\!A_{\alpha }A_{\beta
}^{\ast }A_{\gamma }A_{\delta }^{\ast }\,\mathrm{e}^{\mathrm{i}(S_{\alpha
}-S_{\beta }+S_{\gamma }-S_{\delta })/\hbar }\right\rangle;
\end{eqnarray}%
here $a_{1},c_{1}=1,\ldots ,N_{1}$ and $a_{2},c_{2}=1,\ldots ,N_{2}$
represent ingoing and outgoing channels, connected by the trajectories
$\alpha ,\beta ,\gamma ,\delta $ like $\alpha \left( a_{1}\to
  a_{2}\right)$, $\beta \left( c_{1}\to a_{2}\right)$, $\gamma \left(
  c_{1}\to c_{2}\right)$, $\delta \left( a_{1}\to c_{2}\right)$. The
sum is dominated by quadruplets where the trajectories $\beta $ and
$\delta $ have approximately the same cumulative action as $\alpha $
and $\gamma $, such that the action difference $\Delta S\equiv
S_{\alpha }-S_{\beta }+S_{\gamma }-S_{\delta }$ is of the order $\hbar
$. The contributions of other quadruplets interfere destructively.

\textit{Diagonal contribution:} The simplest quadruplets have either
$\alpha =\beta $, $\gamma =\delta $, or $\alpha =\delta $, $\beta
=\gamma $ \cite{Schanz}; their action difference vanishes.  The
first case has coinciding ingoing channels $a_{1}$ and $c_{1}$ and
contributes
\begin{equation}
\langle \mathrm{tr}(tt^{\dagger }tt^{\dagger })\rangle _{\QATOP{\alpha
=\beta }{\gamma =\delta }}=\frac{1}{T_{H}^{2}}
\sum_{\QATOP{a_{1}}{a_{2},c_{2}
}}\sum_{\QATOP{\alpha \left( a_{1}\to a_{2}\right) }{\gamma \left(
a_{1}\to c_{2}\right) }}|A_{\alpha }|^{2}|A_{\gamma }|^{2}\,.
\label{diag1}
\end{equation}
The foregoing sum can be done using ergodicity. As shown in \cite{RS},
summing all trajectories between two fixed channels amounts to
integrating over the dwell time $T$,
\begin{equation}  \label{sumrule}
\sum_{\alpha(a_1\to a_2)}|A_\alpha|^2=\int_0^\infty dT 
\mathrm{e}^{-\frac{N}
{T_H}T}=\frac{T_H}{N}\,,
\end{equation}
where $\mathrm{e}^{-\frac{N}
{T_H}T}$ is the probability for the
trajectory to dwell in the cavity up to the time $T$, and
$\frac{N}{T_H}$ the rate of escape.

To proceed with Eq.~(\ref{diag1}) we invoke the Richter/Sieber sum
rule (\ref{sumrule}) twice and afterwards sum over all $N_1 N_2^2$
possible combinations of channels with $a_1=c_1$. Similarly, the case
$\alpha=\delta$, $\beta=\gamma$ leads to $N_1^2N_2$ combinations with
coinciding outgoing channels $a_2=c_2$.  Altogether, these so-called
diagonal contributions sum up to
$\frac{N_1^2N_2+N_1N_2^2}{N^2}=\frac{N_1 N_2}{N}$. In the unitary
case, they cancel with $\langle\mathrm{tr}(tt^\dagger)\rangle$ such
that shot noise must be entirely due to different quadruplets of
trajectories.

\begin{figure}[ht]
\begin{center}
  \includegraphics[scale=0.115]{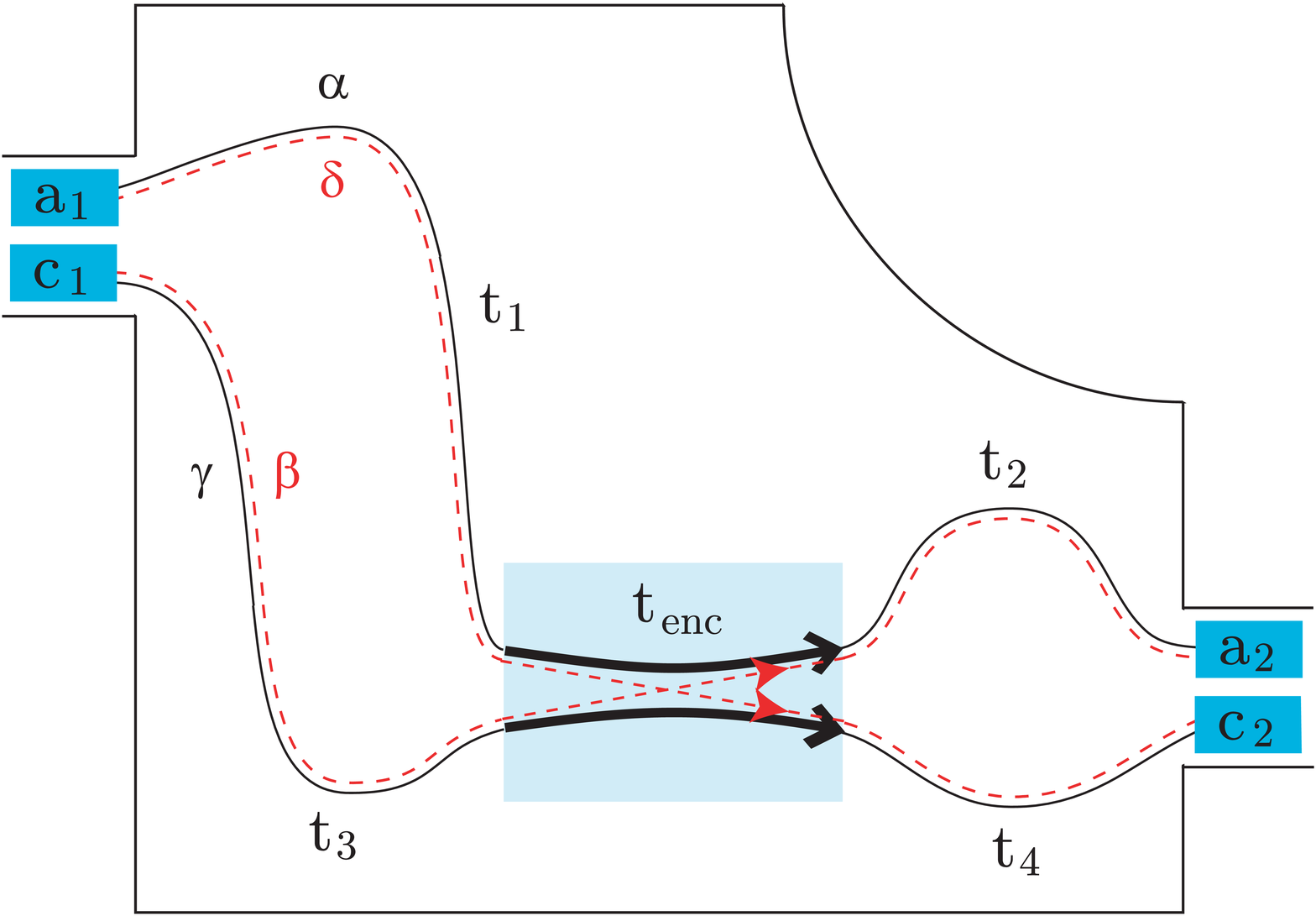}
\end{center}
\caption{Quadruplet of trajectories $\alpha,
  \beta,\gamma,\delta$ differing by their connections inside a
  2-encounter (in the box).  Initial and final points marked by
  channel indices $a_1,c_1,a_2,c_2$.  The durations are $t_{\rm enc}$
  for the encounter and $t_1$, $t_2$, $t_3$, $t_4$ for the loops.}
\label{fig:shotgoett}
\end{figure}

\textit{2-encounters:} The first family of such quadruplets, depicted
in Fig. \ref{fig:shotgoett}, was identified by Schanz, Puhlmann and
Geisel for quantum graphs \cite{Schanz}. %ADD \cite{Lassl}?
Here, the trajectories $\alpha $ and $\gamma$ approach each other in a
``2-encounter": a stretch of $\alpha $ comes so close in phase space
to a stretch of $\gamma $ that the motion over the two stretches is
mutually linearizable.  The remaining parts of $\alpha $ and $\gamma $
will be called \textquotedblleft loops".  Assuming that all loops have
non-vanishing length (Encounter stretches do not "stick out" into the
leads), one finds two further trajectories $\beta $ and $\delta $,
which practically coincide with $\alpha $ and $\gamma $ inside the
loops but are differently connected in the encounter: The trajectory
$\beta $ closely follows the initial loop of $\alpha $ and the final
loop of $\gamma$, whereas $\delta $ follows the initial loop of
$\gamma $ and the final loop of $\alpha $. Obviously, the cumulative
action of $\beta $ and $\delta $ approximately coincides with the
action of $\alpha $ and $\gamma $, with the action difference
exclusively determined by the encounter region.

Each encounter influences the survival probability.
If a particle stays in the cavity along the first encounter
stretch it cannot escape during the second stretch either, since the
two stretches are close to each other. The trajectories $\alpha $ and
$\gamma $ are thus exposed to the danger of getting lost only on the
four loops (see Fig. \ref{fig:shotgoett}) and on \textit{one}
encounter stretch.  Denoting the duration of the latter stretch by
$t_{\rm enc}$ we can write the overall
exposure time as $T_{\mathrm{exp}}=t_{1}+t_{2}+t_{3}+t_{4}+t_{%
  \mathrm{enc}}$. That exposure time is \textit{smaller}
than the cumulative duration $T_{\alpha }+T_{\gamma }$ of $\alpha $
and $\gamma $, by a second summand $t_{\mathrm{enc}}$
for the second encounter stretch. The probability that both
$\alpha $ and $\gamma $
stay inside the cavity reads $\mathrm{e}^{-\frac{N}{T_{H}}T_{\mathrm{%
exp}}}$, larger than the naive estimate $\mathrm{e}^{-\frac{N%
}{T_{H}}(T_{\alpha }+T_{\gamma })}$. In brief, encounters hinder
escape \cite{EssenCond}.

\begin{figure*}
\begin{center}
  \includegraphics[scale=0.2]{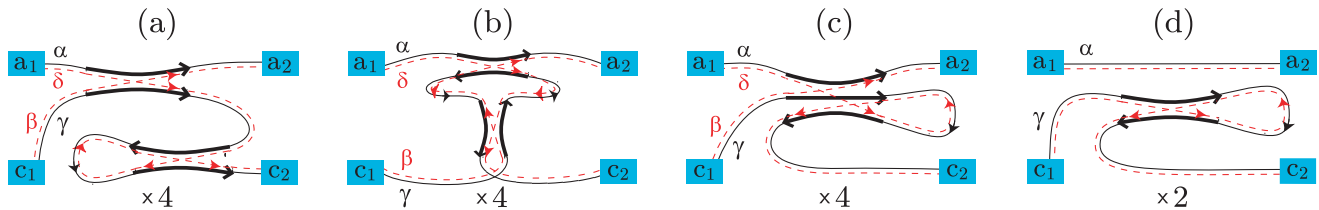}
\end{center}
\caption{Families of trajectory quadruplets 
$\alpha,\beta,\gamma,\delta$
  responsible for the next-to-leading (``weak localization")
  contribution to shot noise (drawn similar as in Fig.
  \ref{fig:shotgoett}, but without the cavity).  Each picture
  represents either four or two similar families of quadruplets.
  Arrows indicate the direction of motion inside the encounters (thick
  lines), and highlight loops which are traversed in opposite
  direction by $(\alpha,\gamma)$ and $(\beta,\delta)$.  }
\label{fig:shotnext}
\end{figure*}

To describe the geometry of encounters, we consider a Poincar\'{e}
section $\mathcal{P}$ in the energy shell, through an arbitrary point
of $\alpha $. If $\mathcal{P}$ cuts through an encounter as in
Fig.~\ref{fig:shotgoett} it must intersect $\gamma $ in a point close
to the reference point on $\alpha $. Assuming two freedoms we
decompose the separation between both points into components $s$, $u$
along the stable and unstable manifolds \cite{EssenFF}.  Both $s$ and
$u$ must be small, $|u|<c,|s|<c$, with $c$ some classically small
constant.  The components $s$ and $u$ fix the action difference as
$\Delta S=su$ and the encounter duration as
$t_{\mathrm{enc}}=\frac{1}{\lambda } \ln \frac{c^{2}}{|su|}$, where
$\lambda $ is the Lyapunov constant \cite{EssenFF}.

Using ergodicity, we count the encounters within trajectory pairs. The
probability density for $\gamma $ to pierce through $\mathcal{P}$ at a
specified time with phase-space separations $s$ and $u$ is uniform and
given by the inverse of the volume of the energy shell $\Omega $. To
capture all encounters, we integrate that density over (i) the time of
piercing on $\gamma $ and (ii) the time of the reference piercing on
$\alpha $.  The probability for $\mathcal{P}$ to cut an encounter is
proportional to the duration $t_{\mathrm{enc}}$, which we divide out
to get the number of encounters \cite{EssenFF}.  Changing the
integration variables to the loop durations $t_{1}$ and $t_{3}$ we
get the weight function
\begin{equation}
w(s,u)=\int_{0}^{T_{\alpha }-t_{\mathrm{enc}}(s,u)}\!\!\!\!\!
dt_{1}\!\!\int_{0}^{T_{%
\gamma }-t_{\mathrm{enc}}(s,u)}\!\!\!\!\!
dt_{3}\frac{1}{\Omega \,t_{\mathrm{enc}%
}(s,u)}\,;  \label{density}
\end{equation}%
here the upper boundaries, depending on the dwell times $T_{\alpha }$,
$T_{\gamma }$ of $\alpha $ and $\gamma $, make sure that the loop
durations $t_{2}$ and $t_{4}$ remain positive. The density $w(s,u)$ is
normalized such that $\int dsdu w(s,u)\delta(su-\Delta S)$ is the
number density of 2-encounters with fixed action difference $\Delta
S$.

To account for all quadruplets of
Fig.~\ref{fig:shotgoett}, we do the sum over $\beta $, $\delta $
in (\ref{quadsum}) by integrating with the weight $w(s,u)$,
\begin{eqnarray}
\!\!\!\!\! &&\langle \mathrm{tr}(tt^{\dagger }tt^{\dagger })\rangle _{%
\mathrm{2-enc}}  \nonumber \\
\!\!\!\!\! &=&\frac{1}{T_{H}^{2}}\left\langle \sum_{\QATOP{a_{1},c_{1}}{%
a_{2},c_{2}}}\int \!ds\,du\sum_{\alpha ,\gamma }|A_{\alpha
}|^{2}|A_{\gamma }|^{2}w(s,u)\mathrm{e}^{{\rm i} su/\hbar }\right\rangle,
\end{eqnarray}%
approximating $A_{\beta }A_{\delta }\approx A_{\alpha
}A_{\gamma }$ \cite{just}.  Now, similarly to
(\ref{sumrule}), we replace the sum over $\alpha $, $\gamma $ by an
integral over the dwell times or, equivalently, over the
loop durations $t_{2}$ and $t_{4}$.  The integrand must be weighted
with the probability $\mathrm{e}^{-\frac{N}{T_{H}}T_{\mathrm{exp}}}$
for both trajectories to remain inside the cavity. Regarding the sum
over channels $a_{1},a_{2},c_{1},c_{2}$ one might expect a factor
$N_{1}^{2}N_{2}^{2}$. However, when both the ingoing channels and the
outgoing channels coincide, $a_{1}=c_{1},\,a_{2}=c_{2}$, each of the
two dashed trajectories in Fig.~\ref{fig:shotgoett} could be chosen
as $\beta $ or $\delta $, such that the resulting contributions are
doubled.  Including such combinations for a second time, we get the
factor $N_{1}N_{2}(N_{1}N_{2}+1)$. Altogether, we thus find
\begin{eqnarray}
\langle \mathrm{tr}(tt^{\dagger }tt^{\dagger })\rangle _{\mathrm{2-enc}}=%
\textstyle{\frac{N_{1}N_{2}(N_{1}N_{2}+1)}{T_{H}^{2}}\big\langle}%
\int\limits_{0}^{\infty
}{dt_{1}dt_{2}dt_{3}dt_{4}}
\nonumber  \label{schanzint} \\
\textstyle{\,\,\times }\int {ds}\,{du\frac{1}{\Omega \,t_{\mathrm{enc}%
}(s,u)}\mathrm{e}^{-\frac{N}{T_{H}}\left[ t_{1}+t_{2}+t_{3}+t_{4}+t_{\mathrm{%
enc}}(s,u)\right] }\mathrm{e}^{\frac{{\rm i} su}{\hbar }}}\big\rangle\,.
\end{eqnarray}%
The integral factors into four independent integrals over the loop
durations, $\int_{0}^{\infty
}dt_{i}\mathrm{e}^{-\frac{N}{T_{H}}t_{i}}=\frac{T_{H}}{N}$, and one
integral over the separations $s$, $u$
inside the encounter, $\int ds\, du\frac{1}{\Omega \,t_{\mathrm{%
 enc}}(s,u)}\mathrm{e}^{-\frac{N}{T_{H}}t_{\mathrm{enc}}(s,u)}\mathrm{e}^{{\rm
    i} su/\hbar }\stackrel{\hbar\to
  0}{\longrightarrow}-\frac{N}{T_{H}^{2}}$ as shown in \cite{EssenCond}.  Since all
powers of $T_{H}$ mutually cancel, the following \textit{diagrammatic
  rule} arises: \textit{Each loop gives rise to a factor
  $\frac{1}{N}$, and an encounter contributes a factor $-N$.} The
rule yields $-1/N^3$; upon multiplying with the number of possible
combinations of channels we get
\begin{equation}
-\langle \mathrm{tr}(tt^{\dagger }tt^{\dagger })
\rangle _{\mathrm{2-enc}}=
\frac{N_{1}N_{2}(N_{1}N_{2}+1)}{N^{3}}\,,
\end{equation}
i.e., for $N_1,N_2\gg 1$ the leading shot-noise term in (\ref{RMT}).

\textit{All orders:} To go beyond the leading term, we have to account
for all quadruplets of trajectories differing in arbitrarily many
encounters, each involving arbitrarily many stretches;
Fig.~\ref{fig:shotnext} shows a few examples. In the unitary case we
must consider encounters where several stretches of either $\alpha$ or
$\gamma$, or both, come close in phase space, whereas in presence of
time-reversal invariance the stretches may also be nearly mutually
time-reversed.

The contributions of all families of quadruplets obey the above
diagrammatic rule. To show this, we describe each $l$-encounter
(encounter of $l$ stretches) by $l-1$ pairs of coordinates $s_{j}$,
$u_{j}$, $j=1,\ldots ,l-1$ \cite{EssenFF} measuring the separations of
$l-1$ stretches from one reference stretch. These coordinates
determine both the duration of each encounter stretch and its
contribution $\sum_{j=1}^{l-1}s_{j}u_{j}$ to the action difference.
The analog of the density $w(s,u)$ in (\ref{density}) obtains a factor
$\frac{1}{\Omega ^{l-1}t_{\mathrm{enc}}(s,u)}$ from each
$l$-encounter.  The resulting product must be integrated over the
durations of all loops, with integration over the final loops of
$\alpha $ and $\gamma $ coming into play through the summation over
$\alpha $ and $\gamma $ like in (\ref{schanzint}). Finally, the
contribution of each family factors into \textquotedblleft loop" and
\textquotedblleft encounter" integrals similar to those in
(\ref{schanzint}). After cancellation of all powers of $T_{H}$, the
diagrammatic rule comes about, with a factor $\frac{1}{N}$ from each
loop and a factor $-N$ from each encounter.

Again, we have to multiply the result with the number of possible
combinations of channels. Two cases must be distinguished.  First, let
us consider trajectory quadruplets as in Fig.~\ref{fig:shotnext}a-c
where similarly to Fig.~\ref{fig:shotgoett} the partner trajectories
$\beta $ and $\delta $ connect the initial point of
$\alpha $ to the final point of $\gamma $, and the initial point of
$\gamma $ to the final point of $\alpha $. Such quadruplets
will be called \textit{$x$-quadruplets}. For them, the channels $%
a_{1}$, $c_{1}$, $a_{2}$, and $c_{2}$ may be chosen arbitrarily and
allow for $N_{1}N_{2}(N_{1}N_{2}+1)$  combinations.

In contrast, Fig. \ref{fig:shotnext}d depicts a quadruplet where the
partner trajectories connect the initial point of $\alpha $ to the
final point of $\alpha$, and the initial point of $\gamma $ to the
final point of $\gamma $ , similarly to the diagonal contribution. We
speak of a \textit{$d$-quadruplet} then. The partner trajectories now
connect the leads as $ a_{1}\to a_{2}$, $c_{1}\to c_{2}$. Since for
shot noise we need partner trajectories $\beta \left( a_{1}\to
  c_{2}\right) ,\delta \left( c_{1}\to a_{2}\right) $,
$d$-quadruplets contribute only if either the two ingoing channels,
or the two outgoing channels (and thus the corresponding angles of
incidence) coincide.  Like for the diagonal contribution, we thus
obtain $N_{1}^{2}N_{2}+N_{1}N_{2}^{2}=NN_{1}N_{2}$ possible
combinations.

Our diagrammatic rules determine $ \langle \mathrm{tr}(tt^{\dagger
}tt^{\dagger })\rangle$ as
\begin{eqnarray}
\langle \mathrm{tr}(tt^{\dagger }tt^{\dagger })\rangle &=&
{\frac{N_{1}N_{2}(N_{1}N_{2}+1)}{N^{2}}}\sum_{m=1}^{\infty }\frac{x_{m}}{N^{m}}
\nonumber  \label{allorders} \\
&&
+{\frac{N_{1}N_{2}}{N}}\left\{ 1+\sum_{m=1}^{\infty }\frac{d_{m}}{N^{m}}%
\right\}\,.
\end{eqnarray}%
Towards explaining $x_m$ and $d_m$ we denote the number
of encounters in a quadruplet by $V$ and the total number of the
encounter stretches by $L$. Then $x_{m}$ is the number of families of
$x$-quadruplets with $m=L-V$ and even $V$, minus the number of
corresponding families with odd $V$; $d_{m}$ is the analogous number
of $d$-families. We note that the contribution of each family is
proportional to $\frac{1}{N^{m+2}}$ rather than $\frac{1}{N^{m}}$,
since there are two more loops than encounter stretches.

The leading contribution to shot noise originates from the family of
Fig.~\ref{fig:shotgoett}; it gives $x_{1}=-1$.  For the
next-to-leading term, we have to consider $x$-quadruplets with two
2-encounters or one 3-encounter, contributing to $x_{2}$ and depicted
in Figs.~2a-c, and $d$-quadruplets which are related to a single
2-encounter and contribute to $d_{1}$ (Fig.~\ref{fig:shotnext}d).  All
quadruplets in Fig.~2 involve mutually time-reversed loops and can
exist in the orthogonal case only (thus $x_{2}=d_{1}=0$ in the unitary
case). Note that if we interchange the two leads, or the pairs
$(\alpha ,\gamma )$ and $(\beta ,\delta )$, or the trajectories
$\alpha $ and $\gamma $, each family of quadruplets will be either
left topologically invariant or turned into an equivalent family
making the same contribution to the shot noise. Only one
representative of each such ``symmetry multiplet" is shown in
Fig.~\ref{fig:shotnext}, with the number of equivalent families
indicated by a multiplier. The sum of all contributions gives
$x_{2}=4,d_{1}=-2$, and $\langle \mathrm{tr} (tt^{\dagger }tt^{\dagger
})\rangle =\frac{N_{1}N_{2}}{N}-\frac{
  N_{1}^{2}N_{2}^{2}}{N^{3}}+4\frac{N_{1}^{2}N_{2}^{2}}{N^{2}}-2\frac{
  N_{1}N_{2}}{N^{2}}+{\cal O}\left( \frac{1}{N}\right) $. Together
with $\langle \mathrm{tr}(tt^{\dagger })\rangle
=\frac{N_{1}N_{2}}{N}-\frac{N_{1}N_{2}}{N^{2}} +{\cal O}\left(
  \frac{1}{N}\right) $, we recover the second term in (\ref{RMT}).

For higher orders in $\frac{1}{N}$, we must collect all families of
trajectory quadruplets. We had previously established a method for
counting families of \textit{pairs} of \textit{ periodic orbits}
differing in encounters, based on permutation theory \cite{EssenFF}.
The families of orbit pairs thus obtained can be turned into the
families of trajectory quadruplets needed now, simply by cutting each
pair twice, inside loops. One can show that if one and one only of the
loops cut is traversed in the opposite sense in the orbits of the
pair, the resulting quadruplet is of type $x$, and otherwise of type
$d$. Using this method we
obtain
\begin{eqnarray}
\label{coeff} x_m&=&\begin{cases}
  \frac{(-1)^m-1}{2}&\text{unitary}\\
  (-1)^m\frac{3^m-1}{2}&\text{orthogonal}
\end{cases}\nonumber\\
 d_m&=&\begin{cases}
  \frac{(-1)^m+1}{2}&\text{unitary}\\
  (-1)^m\frac{3^m+1}{2}&\text{orthogonal}\,.
\end{cases}
\end{eqnarray}
The proof, based on a recursion derived in \cite{EssenFF},
will be given elsewere. Summing over $m$ we get the shot noise 
\begin{equation}\label{univshot}
P=%\langle \mathrm{tr}(tt^{\dagger }-tt^{\dagger }tt^{\dagger })\rangle =%
\begin{cases} \frac{N_1^2 N_2^2}{N(N^2-1)} &\text{unitary}\\
  \frac{N_1(N_1+1)N_2(N_2+1)}{N(N+1)(N+3)}&\text{orthogonal}\,,
\end{cases}
\end{equation}%
valid to all orders in $\frac{1}{N}$ and thus also for a few channels.

\textit{Weak magnetic field:} In the presence of a weak magnetic field
$B$, the power of shot noise must interpolate between the orthogonal
and unitary cases. A weak field
increases the action of each trajectory \cite{NagaoTau3,EssenCond} by
the line integral $\frac{1}{e}\int \mathbf{A}\cdot d\mathbf{q}$ of the
vector potential $\mathbf{A}$. Since that increment changes sign under
time reversal a net contribution survives from all loops and encounter
stretches changing direction in $(\beta ,\delta )$
relative to $(\alpha ,\gamma)$.
Our above diagrammatic rule is thus modified: {\it Each loop changing
direction contributes $\frac{1}{N(1+\xi )}$ with $\xi \propto B^{2}$;
each encounter contributes $-N(1+\xi \mu ^{2})$ with $\mu$ the number
of its stretches changing direction} \cite{EssenCond}.

The leading contribution to shot noise, from quadruplets as in
Fig.~\ref{fig:shotgoett}, remains unaffected by the magnetic field.
However, all quadruplets responsible for the next-to-leading term
obtain a Lorentzian factor $\frac{1}{1+\xi }$. (In
Figs.~\ref{fig:shotnext}a, \ref{fig:shotnext}c, and
\ref{fig:shotnext}d one loop is traversed in time-reversed sense and
$\mu=0$, whereas in Fig.~\ref{fig:shotnext}b two loops are
time-reversed and one encounter has $\mu =1$.)  We thus predict
\begin{equation}
\label{Pmagn}
P=\frac{N_{1}^{2}N_{2}^{2}}{N^{3}}+\frac{N_{1}N_{2}(N_{1}-N_{2})^{2}}
{N^{4}(1+\xi )}+{\cal O}\Big( \frac{1}{N}\Big) \,,
\end{equation}
in accordance with (\ref{univshot}) for the limits $\xi \to 0$ and
$\xi \to \infty $. If $N_1=N_2=N/2$ the second term in (\ref{Pmagn})
vanishes, and quadruplets involving more encounter stretches yield
\begin{equation}
P=\frac{N}{16}+\frac{1}{N}\frac{1+8\xi+4\xi^2+4\xi^3+\xi^4}{%
16(1+\xi )^4}+{\cal O}\Big( \frac{1}{N^2}\Big) \,;
\end{equation}%
the extension to ${\cal O}(\frac{1}{N^{6}})$ will be given elsewhere.

{\it Outlook:} The semiclassical approach to transport opens a large
field of experimentally relevant applications. For instance, we have
checked the trajectory quadruplets employed here to also yield the
universal conductance variance as well as the covariance of the
conductance at two different energies, both within and in between
the orthogonal and unitary symmetry classes.

We are indebted to Dmitry Savin, Hans-J{\"u}rgen Sommers (who have
reproduced our prediction (\ref{univshot}) in random-matrix theory),
and Martin Zirnbauer for useful discussions and to the
Sonderforschungsbereich SFB/TR12 of the Deutsche
Forschungsgemeinschaft for financial support.


\begin{thebibliography}{99}
\bibitem{just} This approximation is justified because the stability
  amplitudes of $\alpha $ and $\gamma $ can be approximated as
  products of contributions from loops and stretches, practically
  coinciding with those of $\beta $ and $\delta $.
\end{thebibliography}
\end{document}